\documentclass[conference]{IEEEtran}
\IEEEoverridecommandlockouts

\usepackage{url}
\usepackage{float}
\usepackage{svg}
\usepackage[utf8]{inputenc}
\usepackage{pdfpages}
\usepackage{graphicx}
\usepackage{wrapfig}
\usepackage{multicol}
\usepackage{svg}
\usepackage{MnSymbol}
\usepackage{hyperref}
\usepackage{amsfonts} 
\usepackage{booktabs}
\usepackage{cite}
\usepackage{amsmath,amsfonts}
\usepackage{algorithmic}
\usepackage{textcomp}
\usepackage{xcolor}
\usepackage{tikz}
\usetikzlibrary{fit, shapes.geometric, backgrounds}
\usepackage{pgfplots}
\usepackage{pifont}
\definecolor{color2}{RGB}{154, 205, 50}
\newcommand{\cmark}{\ding{51}}  
\newcommand{\xmark}{\ding{55}}
\newcommand{\greenstar}{
  \begin{tikzpicture}[baseline=-0.5ex]
    \node[star, star points=5, star point ratio=2.25, fill=green, scale=0.5] {};
  \end{tikzpicture}
}
\def\BibTeX{{\rm B\kern-.05em{\sc i\kern-.025em b}\kern-.08em
    T\kern-.1667em\lower.7ex\hbox{E}\kern-.125emX}}
    
\begin{document}

\title{LHGNN: Local-Higher Order Graph Neural Networks For Audio Classification and Tagging\\

\thanks{S. Singh was supported by UKRI and Queen Mary University of London under Grant EP/S022694/1. E. Benetos was supported by RAEng/Leverhulme Trust Research Fellowship LTRF2223-19-106. This work is not related to H. Phan's work at Meta.}
}

\author{
\IEEEauthorblockN{Shubhr Singh\textsuperscript{1}, Emmanouil Benetos\textsuperscript{1}, Huy Phan\textsuperscript{2}, and Dan Stowell\textsuperscript{3}}
\IEEEauthorblockA{
    \textsuperscript{1}School of Electronic Engineering and Computer Science, Queen Mary University of London, UK \\
    \textsuperscript{2}Meta, 75002 Paris, France \\
    \textsuperscript{3}Tilburg University, Bijsterveldenlaan, 5037 AB Tilburg, Netherlands
}
}


\maketitle

\begin{abstract}
Transformers have set new benchmarks in audio processing tasks, 
%
leveraging self-attention mechanisms to capture complex patterns and dependencies within audio data. However, their focus on pairwise interactions limits their ability to process the higher-order relations essential for identifying distinct audio objects. To address this limitation, this work introduces the Local-Higher Order Graph Neural Network (LHGNN), a graph based model that enhances feature understanding by integrating local neighbourhood information with higher-order data from Fuzzy C-Means clusters, thereby capturing a broader spectrum of audio relationships. Evaluation of the model on three  publicly available audio datasets
shows that it outperforms Transformer-based models across all benchmarks while operating with substantially fewer parameters. Moreover, LHGNN demonstrates a distinct advantage in scenarios lacking ImageNet pretraining, establishing its effectiveness and efficiency in environments where extensive pretraining data is unavailable.
\end{abstract}


\section{Introduction}
The realm of audio classification and tagging has evolved rapidly with the adoption of deep learning technologies. Spanning sound event detection~\cite{mesaros2021sound} to advanced applications like music recommendation~\cite{schedl2019deep} and keyword spotting~\cite{lopez2021deep}, the impact of these technologies is profound.
Historically, CNNs were the preferred architecture for audio classification~\cite{palanisamy2020rethinking} until Transformers~\cite{gong2021ast} demonstrated their superiority in handling complex interactions and larger datasets. While convolutional layers use learnable kernels that reduce overfitting and enhance generalization (especially beneficial with smaller datasets due to their strong inductive bias), Transformers, with their adaptive attention mechanism, excel in modeling more intricate patterns by mapping a global receptive field from the first layer itself. 

Another compelling line of research in deep learning architectures explores the integration of clustering methods with Transformers for tasks such as image classification and object detection~\cite{liang2024clusterfomer}. The process involves projecting features into a set of cluster centers and subsequently redistributing these cluster centers back into the original feature space using similarity metrics. This approach conceptually mirrors the operations of a specialized form of Graph Neural Network (GNNs) known as Hypergraph Neural Networks (HGNNs)~\cite{feng2019hypergraph,han2023vision}. In HGNNs, node features are first projected onto hyperedges, and then updated node features are obtained by projecting back from these hyperedges. Although in deep learning literature, parallels have been drawn between transformers and GNNs~\cite{velivckovic2023everything}, positioning transformers as a specialized iteration of the latter, only recently have graph neural networks been employed in vision~\cite{han2022vision} and audio~\cite{singh2024atgnn}.\vspace{0.1cm}

In this work, we introduce Local-Higher Order Graph Neural Networks (LHGNN), a model which integrates the robust capabilities of GNNs with clustering techniques. LHGNN utilizes local relationships through the k-nearest neighbor (k-NN) algorithm and higher-order relationships via Fuzzy C-Means clustering, enhancing the model by transcending the pairwise interactions typical in standard Transformers and graph-based methods.
Fuzzy C-Means \cite{han2023vision} extends traditional k-means by allowing probabilistic cluster assignments, enabling data points to belong to multiple clusters with varying degrees of membership. Integrating local neighborhood information with higher-order clustering in our LHGNN model offers two benefits: (i) it enables the modeling of higher-order semantic relationships by leveraging clustering techniques, and (ii) it facilitates the modeling of multi-scale relationships in audio by integrating local k-NN and higher-order clustering information.\vspace{0.1cm}

The key contributions of this paper are: (i) the introduction of a novel graph kernel for graph neural networks that integrates local and higher-order interactions for robust representations, and (ii) demonstration of the model's robust performance without the need for extensive ImageNet pretraining, enhancing its versatility in both data-rich and data-scarce environments.

\begin{figure*}[t]
  \centering
  \includegraphics[width=0.9\textwidth]{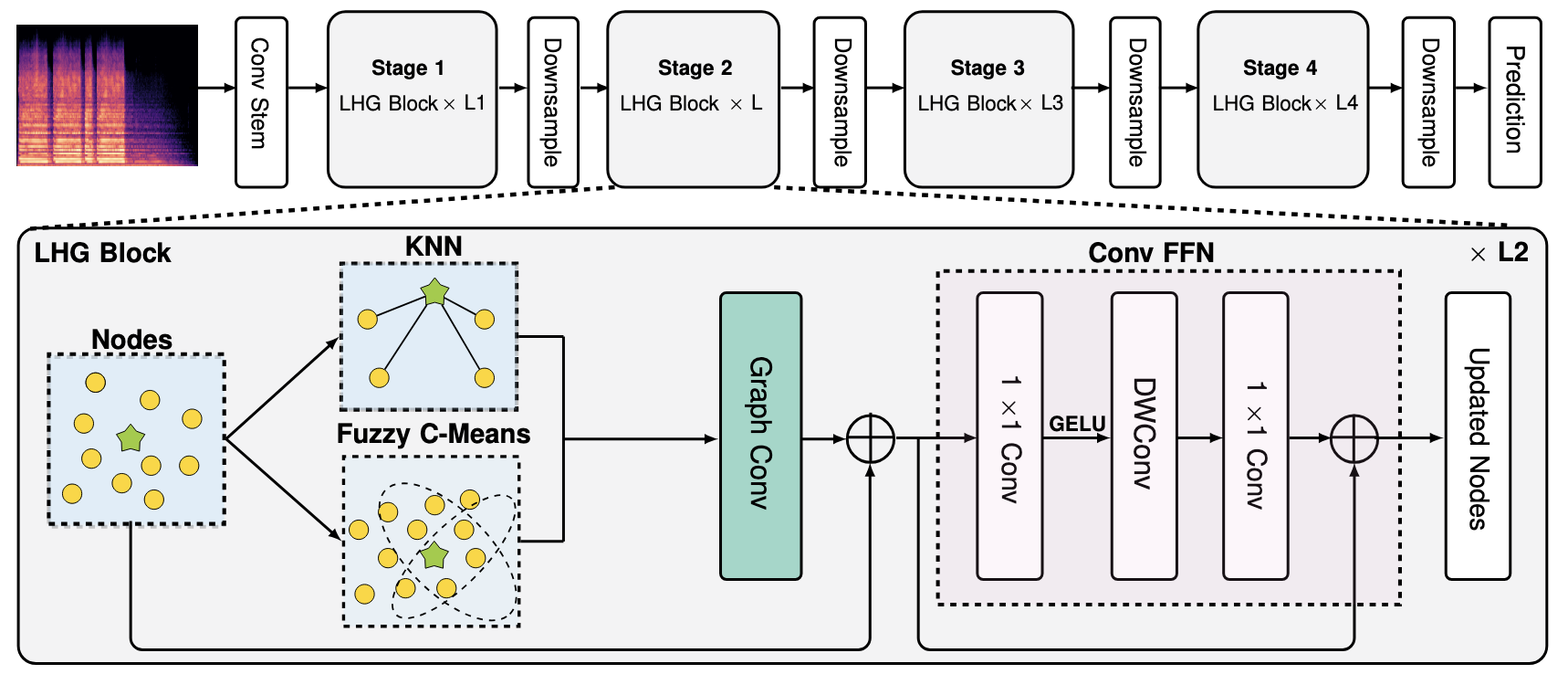}
\caption{\label{fig:figure1}\textbf{Architecture of LHGNN}: Input mel-spectrogram is processed through a convolution block and sent to LHG blocks. In each of the LHG blocks,  \protect\greenstar\ (a single node)  is updated through first constructing a k-NN graph and simulataneously conducting Fuzzy C-Means. The local (k-NN graph) and higher order (cluster centers from Fuzzy C-Means) are fused together to update \protect\greenstar\ , followed by a graph convolution and subsequently sent to  ConvFFN block. DWConv in the ConvFFN block refers to Depthwise Convolution. $L$ represents the number of repetitions for the LHG blocks.}
\end{figure*}

\section{Method}
\label{sec:method}

\subsection{Model Architecture}

A high level overview of the model architecture is illustrated in Fig ~\ref{fig:figure1}. The input mel-spectrogram is first processed through a stem block that consists of  four $3\times 3$ convolutional layers with strides of $2$, $1$, $2$, and $1$ respectively. In contrast to the traditional non-overlapping tokenization approach, the convolution backbone is capable of extracting superior local representations and has become widely adopted in modern Vision Transformers (ViTs)~\cite{guo2022cmt}. The resulting feature map 
is fed into four stages of the stacked Local-Higher Order Graph (LHG) blocks. 

Between the stages of the network, downsampling blocks that include $3\times 3$ convolutions with a stride of $2$ are employed to decrease the number of tokens. The output from the final downsampling block undergoes global average pooling, followed by a $1 \times 1$ convolution and a fully connected layer to produce the final predictions.    

\subsection{LHG Block}

The LHG block consists of two main components: Local-Higher Order Graph Convolution  and Convolutional Feed Forward Network (ConvFFN). 

\subsubsection{Local-Higher Order Graph Convolution}
\label{subsec:lhgnn}
The output from the convolutional backbone is denoted as $\mathcal{X}$, which is a feature map with dimensions \( \mathbb{R}^{H \times W \times C} \). Here, \( H \), \( W \), and \( C \) represent the height, width, and number of channels, respectively. To prepare this data for subsequent analysis, we initially flatten the feature map
to obtain a set of nodes $\mathcal{X} = \{x_1, x_2, \ldots, x_N\} \in \mathbb{R}^{N \times C}$, (where $N = H \times W $). For each node $x_i$, we perform the following simultaneous operations:
    
    \textbf{(i) k-NN} -  Identify the \( k \) nearest neighbors of $x_i$, forming a local subset $\mathcal{S}_i \subset X$. This can be expressed as:
\[ \mathcal{S}_i = \operatorname{k-NN}(x_i, \mathcal{X}, k) \]

    \textbf{(ii) Fuzzy C-Means Clustering} - Apply Fuzzy C-Means clustering to obtain membership scores for \( x_i \) relative to \( P \) centroids.  
    The membership score \( u_{ip} \) of a data point \( x_i \) to the \( p \)-th centroid, \( c_p \), is defined as:
\begin{equation}
u_{ip} = \frac{1}{\sum_{j=1}^P \left(\frac{d(x_i, c_p)}{d(x_i, c_j)}\right)^{\frac{2}{m-1}}}
\end{equation}
where \( d(x_i, c_j) \) represents the Euclidean distance between \( x_i \) and centroid \( c_j \), and \( m \) is the fuzziness parameter that controls the degree of fuzziness in the clustering. The  parameter $m$ is commonly set to 2 in Fuzzy C-Means clustering, as this is a widely accepted standard. Accordingly, we adhere to this typical value for $m$ in all our experiments.
    
Once the membership scores are computed, the centroids are  updated in the following manner:
\begin{equation}
c_p = \frac{\sum_{i=1}^N u_{ip}^m x_i}{\sum_{i=1}^N u_{ip}^m}
\end{equation}
The entire process of calculating membership scores and centroid updates repeats for $v$ iterations. Although higher value of $v$ results in more robust centroids, it consumes significant amount of time even for small number of centroids, hence we restrict $v =1$. 


The set of K centroids with highest \( u_{ip}^m \) are then selected to form the set \( \mathcal{L}_i \) for the data point \( x_i \).


Given $\mathcal{S}_i$ and $\mathcal{L}_i$, we update node $x_i$ through the proposed graph convolution in the following manner:
\begin{equation}
    x_i^{''} = \sigma (x_i \oplus \max(\mathcal{S}_i - x_i) \oplus \max(\mathcal{L}_i -x_i))
\end{equation}
where $\sigma$ denotes a non-linear operation implemented by an MLP network with GELU ~\cite{hendrycks2016gaussian} non-linearity and $ \oplus$ denotes concatenation operation.  The proposed graph convolution, a variant of the max-relative graph convolution \cite{han2022vision}, is specifically designed to capture hierarchical and multiscale relationships. The operation $\max(\mathcal{S}_i - x_i)$ involves first subtracting the central node $x_i$ from each node in the set $\mathcal{S}_i$ on an element-wise basis. Then, the max operation is applied across the resulting differences to capture the maximum deviation of the neighboring nodes from the central node along each feature dimension. Similarly, the operation $\max(\mathcal{L}_i - x_i)$ follows the same methodology but on a broader scale. 
 $x_i'' \in \mathbb{R}^{1 \times 3C}$ is mapped back to the original dimensionality of \(x_i\) using a linear projection function \(h(\cdot)\), and then added to \(x_i\) to produce the final updated node \(y_i\):
\begin{equation}
    y_i = x_i + h(x_i'').
\end{equation}

\subsubsection{ConvFFN}
\label{subsec:convffn}
ConvFFN is applied to each updated node embedding that emerges from the local-higher order graph convolution.
A ConvFFN block, as proposed by ~\cite{huang2022orthogonal},
consists of two $1\!\times\!1$ convolutions, one $3\!\times\!3$ depth-wise
convolution and one non-linear function, i.e., GELU. While Feed-Forward Networks (FFNs) were originally introduced within the context of Transformers, characterized by two linear layers separated by a non-linear activation, the incorporation of depthwise convolution serves to preserve local information across layer depths. 

Notably, prior research indicates that self-attention acts like a low-pass filter~\cite{park2022vision} and ConvFFN counteracts this effect by preserving high-frequency information~\cite{zhou2023srformer}, hence  we employ this block to retain local correlation information throughout the layers.

\subsubsection{Downsample Block}
The ConvFFN block output is reshaped to $\mathbb{R}^{H \times W \times C}$ and then processed by a downsampling block, reducing dimensions by a factor of $r$ to $\mathbb{R}^{\frac{H}{r}\times \frac{W}{r} \times C^t}$, where $r$ is the downsampling ratio and $C^t$ is the new channel count at stage $t$. Downsampling is achieved by applying a \texttt{Conv2d} layer with a \(3 \times 3\) kernel, stride 2, and padding 1.
The downsampled feature map  serves as input for the next stage, repeating processes from Sections~\ref{subsec:lhgnn} and~\ref{subsec:convffn}.

\subsection{Implementation \& Pretraining Details}

We follow a pyramid architecture similar to \cite{han2022vision}, where the channel dimensions progressively increase within each block, following the sequence $[80, 160, 320, 640]$. The LHG blocks are iteratively applied, repeated in the sequence of $[2, 2, 6, 2]$ for stages 1, 2, 3, and 4, respectively. Our best results are obtained with $k = 25$ for k-NN and $K=10$ for selecting the top $K$ centroids based on membership scores. 
The number of centroids $P$ remains constant at $50$ across all stages and for ImageNet pretraining, we adapted the training protocol from~\cite{han2022vision}, modifying the batch size to 512 and reducing the learning rate to $1e-3$. Also, due to input size mismatch, the best results are obtained with $k = 9$ for k-NN and $K=5$ for ImageNet pretraining.

\section{Experiments}
\label{sec:experiments}

We assess the model's performance across two tasks: tagging and classification. Audio tagging evaluation is conducted on Audioset~\cite{AUDIOSET} and FSD50K~\cite{fsd50k}. For audio classification, the model is evaluated using the ESC50 dataset~\cite{ESC50}.

\subsection{Audioset Experiments}
\subsubsection{Dataset and Experimental Procedure}
AudioSet~\cite{AUDIOSET} comprises over 2 million 10-second audio clips extracted from YouTube videos, categorized into 527 sound event classes. It is a weakly labeled and multi-labeled dataset, where each clip can have various tags, but specific timestamps for the onset and offset of these labels are not provided.

We trained our model on the full-train set (2M samples) and evaluated it on the evaluation set (22K samples). All audio samples were converted to mono with a sampling rate of 16kHz. We computed the Short-time Fourier transform (STFT) using a window size of 25 ms and a hop size of 10 ms. A 128-dimensional mel filter bank was applied, followed by a logarithmic transformation to extract the log-mel spectrogram. To ensure uniformity, we standardized the temporal length of the mel-spectrogram to 1024 frames, resulting in a consistent shape of (1024, 128). Shorter clips were zero-padded, and longer ones cropped.

Following the training pipeline suggested in~\cite{PSLA}, we used mixup~\cite{zhang2017mixup} data augmentation with  $\alpha = 0.5$, spectrogram masking~\cite{specaugment} with a time-mask of $192$ frames and frequency mask of $48$ bins. 
The LHGNN was implemented in PyTorch and trained using the AdamW optimizer with parameters \(\beta_1 = 0.9\), \(\beta_2 = 0.999\), \(\epsilon = 10^{-8}\), and a decay rate of 0.05. Training was conducted with a batch size of 128, distributed across four NVIDIA Tesla A100 GPUs.

\subsubsection{Results on Audioset}

In Table~\ref{tab:table1}, we compare our model with different benchmark models.  DeepRes~\cite{ford2019deep}, PANN~\cite{kong2020panns} and PSLA ~\cite{PSLA} are CNN based models and AST~\cite{gong2021ast} is a transformer based model. The reported scores for AST, PSLA, and LHGNN were calculated using  weighted average of different model checkpoints as mentioned in ~\cite{PSLA}. Notably, the LHGNN model surpasses AST in performance while utilizing a significantly smaller number of parameters. A key observation is the distinct performance gap between AST and LHGNN when neither model is pretrained with ImageNet. This underscores the significant influence of ImageNet pretraining on supervised audio based tasks. The impact of such pretraining is further highlighted by comparing the performance outcomes of models like DeepRes, which lacks ImageNet training, to those that include it, such as PSLA, and AST. ImageNet pretraining is resource-intensive and time-consuming. However, LHGNN performs exceptionally well without pretraining, demonstrating the model's robustness and efficiency. 

\begin{table}[]
    \centering
    \caption{Results on AudioSet}
    \begin{tabular}{l c c c} \toprule
        \textbf{Model} & \textbf{\#Params} &
        \textbf{Pretrain} &
        \textbf{mAP} \\ \midrule
        Baseline~\cite{AUDIOSET}  & 2.6\,M & \xmark  & 0.314\\
        DeepRes~\cite{ford2019deep}  & 26\,M & \xmark  & 0.392\\
        PANN~\cite{kong2020panns}  & 81\,M & \xmark  & 0.434\\
        
        PSLA~\cite{PSLA}  & 13.6\,M & \cmark  & 0.444\\
        
        AST~\cite{gong2021ast}  & 87\,M & \xmark  & 0.366\\

        AST~\cite{gong2021ast}  & 87\,M & \cmark  & 0.459\\
        LHGNN  & 31\,M & \xmark  & 0.442\\
        LHGNN  & 31\,M & \cmark  & \textbf{0.466}\\
          \bottomrule
    \end{tabular}
    \label{tab:table1}
\end{table}

\subsection{FSD50K Experiments}
\begin{table}[]
    \centering
    \caption{Results on FSD50K}
    \begin{tabular}{l c c c} \toprule
         \textbf{Model} & \textbf{\#Params} &
         \textbf{Pretrain} &
          \textbf{ mAP}  \\\midrule
        
         FSD50K Baseline~\cite{fsd50k}  & 0.27M  & \xmark & 0.434\\
        
        Wav2CLIP~\cite{wu2022wav2clip} & - & \xmark &  0.431 \\
        
        Audio Transformers~\cite{verma2021audio} & 2.3M & - &  0.537\\
        
        PSLA~\cite{PSLA}  &  13.6M & \cmark & 0.559 \\

        AST~\cite{gong2021ast} & 87M & \xmark &  0.396 \\ 
         AST~\cite{gong2021ast} & 87M & \cmark &  0.574 \\

         LHGNN& 31M & \xmark &  \textbf{0.573} \\

         LHGNN& 31M & \cmark &  \textbf{0.59} \\

          \bottomrule
    \end{tabular}
    \label{tab:table2}
\end{table}
\subsubsection{Dataset and Experimental Procedure}

FSD50K~\cite{fsd50k} is a public dataset of weakly labeled sound event audio clips, classified into 200 categories using the AudioSet ontology. It consists of 37,134 training samples, 4,170 validation samples, and 10,231 evaluation samples. Like AudioSet, FSD50K is multi-labeled. We applied the same feature extraction and data augmentation pipeline as in the AudioSet experiments.

\subsubsection{Results on FSD50K}
In Table~\ref{tab:table2}, we compare our model with different benchmark models. FSD50K baseline is a CNN based model, whereas Wav2CLIP~\cite{wu2022wav2clip} employs distillation from contrastive language-image  pre-training (CLIP)~\cite{radford2021learning}. Audio Transformer, like AST, is a self-attention model but uses a learnable MLP frontend to extract representations directly from raw audio.
As shown in Table~\ref{tab:table2}, LHGNN with ImageNet pretraining achieves the best score compared to the benchmark models. Additionally, when trained from scratch, it delivers results comparable to AST with ImageNet pretraining, demonstrating its effectiveness even without relying on large-scale pretraining.

\subsection{ESC50 Experiments}
\begin{table}[t]
    \centering
    \caption{Results on ESC50}
    \begin{tabular}{lccc} \toprule
         \textbf{Model} & \textbf{\#Params} &
         \textbf{Pretrain} &
          \textbf{ Accuracy(\%)}  \\\midrule
        
         PANN~\cite{kong2020panns}  & 81M  & \cmark & $94.7$ \\
        
        AST~\cite{gong2021ast} & 87M & \cmark & $95.6 \pm 0.4$ \\

        ERANN~\cite{verbitskiy2022eranns} & 38.2M & \xmark &  $96.1$ \\
         LHGNN& 31M & \cmark &  {$ 96.2 \pm 0.3$} \\

          \bottomrule
    \end{tabular}
    \label{tab:table3}
\end{table}

\subsubsection{Dataset and Experimental Procedure} ESC50~\cite{piczak2015dataset} is a multi-class audio dataset consisting of $2000$ audio clips, each with 5-sec duration. It is labelled with 50 environmental sound classes across 5 folds. Our model was trained for 5 times by selecting 4-folds (1600 samples) as training and 1-fold (400 samples) as test set. The entire experiment was repeated for 5 times with different random seeds to get the mean score along with its deviation. Accuracy is used as the evaluation metric for all experiments.

\subsubsection{Results on ESC50}

We evaluate the ImageNet trained LHGNN on ESC50 dataset  and observe that the model performs well on multi-class scenario as well. However, as shown in Table~\ref{tab:table3}, the ERANN~\cite{verbitskiy2022eranns} model performs equally well without pretraining. 

\section{Ablation Study}
We conducted ablation studies on the FSD50K dataset without pretraining to optimize our model's parameters. FSD50K was chosen for its balance between size and scalability.


\subsubsection{Graph Kernel}

\begin{table}[]
    \centering
    \caption{Results on FSD50K with different kernels}
    \begin{tabular}{lc} \toprule
         \textbf{Kernel}
          & \textbf{mAP}  \\\midrule
        
         $(x_i \oplus \max(\mathcal{S}_i - x_i))$  & 0.531  \\
        
         $(x_i \oplus \max(\mathcal{L}_i - x_i))$  & 0.501  \\

         $(x_i \oplus \max(\mathcal{S}_i - x_i) \oplus \max(\mathcal{L}_i - x_i))$  & \textbf{0.573}  \\

          \bottomrule
    \end{tabular}
    \label{tab:table4}
\end{table}

As shown in Table ~\ref{tab:table4}, combining local feature information with cluster centroids produced the best results, likely due to the loss of local information when solely employing cluster information in $(x_i \oplus \max(\mathcal{L}_i - x_i))$.


\subsubsection{Clustering Method}
Table ~\ref{tab:table5} compares k-means, Fuzzy C-Means, and density-based clustering. While density-based clustering slightly outperformed Fuzzy C-Means, the latter was chosen for its computational efficiency.

\begin{table}[t]
    \centering
    \caption{Clustering method evaluation on FSD50K}
    \begin{tabular}{lc} \toprule
         \textbf{Clustering method}
          & \textbf{mAP}  \\\midrule
        
         k-means  & 0.544  \\
        
         Fuzzy C-Means  & 0.573  \\

         Density based clustering  & \textbf{0.574}  \\

          \bottomrule
    \end{tabular}
    \label{tab:table5}
\end{table}

\section{Discussion and Conclusion}

This paper presents LHGNN, a new model that combines graph neural networks with clustering techniques to improve audio classification and tagging. Our experiments showed that LHGNN outperforms AST models across multiple datasets, including Audioset, FSD50K, and ESC-50, performing notably well even without pretrained weights.

Its key innovation in the proposed model is the combination of k-nearest neighbor graphs and Fuzzy C-Means clustering to capture complex audio patterns. Despite strong performance, LHGNN takes longer to converge, requiring 30 epochs on Audioset compared to 5 for AST with ImageNet pretraining. Furthermore, a more efficient method for integrating cluster centroids and local information needs to be devised in order to reduce the overall computation time. Ultimately, evaluating the model's performance across a spectrum of audio tasks, such as music tagging and speech recognition, becomes imperative to affirm its efficacy and versatility. This approach is particularly essential given the demonstrated success of Transformers across a diverse range of audio applications. We intend to address these limitations in our future work.



In conclusion, LHGNN stands as a significant step forward in the field of audio classification and tagging, providing a robust framework that leverages graph-based and clustering methodologies to achieve high performance. 

\bibliographystyle{IEEEbib}
\bibliography{strings,references}

\end{document}